\documentclass[]{interact}

\usepackage{epstopdf}
\usepackage[caption=false]{subfig}

\usepackage{multirow}
\usepackage{rotating}
\usepackage{enumerate}
\usepackage{csquotes}
 \usepackage{float} 
 \usepackage{appendix}
 \usepackage{array} 
 \usepackage{hyperref}
 \hypersetup{
    colorlinks=true,
    linkcolor=blue,
    filecolor=magenta,      
    urlcolor=blue,
    citecolor=blue
    }

\usepackage[natbibapa,nodoi]{apacite}
\DeclareMathOperator*{\argmax}{arg\,max}

\theoremstyle{plain}

\theoremstyle{definition}

\theoremstyle{remark}

\begin{document}

\title{Dynamic groups in complex task environments: To change or not to change a winning team?}

\author{
\name{Dari­o Blanco-Fernandez, Stephan Leitner\thanks{CONTACT Stephan Leitner. Email: stephan.leitner@aau.at}, 
Alexandra Rausch}
\affil{University of Klagenfurt, Universit\"atsstra{\ss}e 65-67, 9020 Klagenfurt, Austria}
}

\maketitle

\begin{abstract}
Organisations rely upon group formation to solve complex tasks, and groups often adapt to the demands of the task they face by changing their composition periodically. Previous research comes to ambiguous results regarding the effects of group adaptation on task performance.
This paper aims to understand the impact of group adaptation, defined as a process of periodically changing a group's composition, on complex task performance and considers the moderating role of individual learning and task complexity in this relationship. We base our analyses on an agent-based model of adaptive groups in a complex task environment based on the \textit{NK} framework. The results indicate that reorganising well-performing groups might be beneficial, but only if individual learning is restricted. However, there are also cases in which group adaptation might unfold adverse effects. We provide extensive analyses that shed additional light on and, thereby, help explain the ambiguous results of previous research. 
\end{abstract}


\section{Introduction}\label{sec:intro}
Stability in group composition is often considered a desirable trait. For example, Alf Ramsey led the England National Football Team to victory in the 1966 World Cup by \textit{never changing a winning team}. In contrast, \cite{Nuesch2012} find that managers in the German Bundesliga who apply this strategy do not significantly improve their team’s performance. Nevertheless, managers use all available in-game substitutions during a match \citep{Myers2012}, and squad changes during knockout tournaments are a standard \citep{Mengel2009}. Decisions about changes in the squad, both in- and out-game, depend on factors such as a result during the game \citep{Myers2012}, position in the tournament's first-round \citep{Mengel2009}, and injury prevention \citep{Varela-Quintana2016}. Thus, managers seem to change the team composition to \textit{adapt} to the current conditions of the game. This paper takes up this idea and examines the effects of \textit{group adaptation} to the task environment on performance \citep{Bell2017,Tannenbaum2012}.

In Organisational Science, there are three main strands of related research. First, conceptual research is concerned with the temporal aspects of groups, including the emergence of efficient group compositions \citep{Bell2017,Lundin1995,Tannenbaum2012}. Second, there is research on the relationship between group adaptation and learning \citep{Bartsch2013,Edmondson2001,Edmondson2003,Savelsbergh2015,Sergeeva2018}. 
Third, some research focuses on the effects of group adaptation on creativity and innovation \citep{Choi2004,Choi2005,Perretti2007,Spanuth2020}. However, research often lacks directly linking group adaptation to task performance. 

We place our research in this gap and consider task complexity and individual learning as moderating factors in the relationship between group adaptation and performance. Previous research has extensively shown that task complexity --in terms of the number and pattern of the interdependencies between subtasks (i.e., the actions that need to be taken to perform a task)-- highly affects performance \citep{Levinthal1997,Rivkin2000,leitner14,Leitner2021}. Concerning learning, \cite{Simon1991} argues that there might be an interaction between individual learning (i.e., creating knowledge within a group) and changing a group's composition (i.e., absorbing knowledge outside of the group by attracting new members), since--from the group's perspective--both approaches result in new knowledge. Consequently, we expect moderating effects of learning and task complexity in the relationship between group adaptation and performance. 

By studying the effect of group adaptation on task performance, we attempt to determine whether \textit{never changing a winning team} is a sustainable strategy. Taking into account all factors, we formulate the following three research questions:

\begin{enumerate}[(i)]
    \item What is the effect of group adaptation on task performance?
    \item How does individual learning moderate the relationship between group adaptation and task performance?
    \item How does complexity moderate the relationship between group adaptation and task performance?
\end{enumerate}

\noindent To answer these questions, we propose an agent-based model based on the \textit{NK} framework \citep{Levinthal1997,Leitner2015,Wall2020}. The remainder of this paper is organised as follows: We provide the background for group adaptation in Sec. \ref{sec:background}. Section \ref{sec:model} introduces the agent-based model. The results and a discussion are presented in Sec. \ref{sec:results}. Finally, Sec. \ref{sec:conclusion} concludes the paper.

\section{Group adaptation}\label{sec:background}

According to \cite{Bell2017}, \cite{Tannenbaum2012}, and \cite{Lundin1995}, organisations increasingly rely on \textit{team}-based structures for their operations, 
whereby temporariness is a crucial aspect of these groups. \cite{Lundin1995} claim that groups usually change their composition after completing a task. Following this argument, \cite{Bell2017} and \cite{Tannenbaum2012} state that these changes repeatedly occur in response to the task's demands. 

According to prior research, changes in group composition might result in \textit{(i)} increasing creativity and the diversity of solutions employed to solve a task \citep{Choi2004,Choi2005,Perretti2007}, 
\textit{(ii)} integrating the best-available experts within the group's ranks \citep{Savelsbergh2015}, and 
\textit{(iii)} offsetting of the positive effects that individual learning has on task performance \citep{Bartsch2013,Sergeeva2018}. Consequently, this line of research argues that any decision involving group adaptation results in a trade-off, and groups should balance the advantages (i.e., \textit{(i)} and \textit{(ii)}) and disadvantages (i.e., \textit{(iii)}) of adaptive groups \citep{Savelsbergh2015}. Thus, these results call the strategy to \enquote{never change a winning team} into question.

However, how group adaptation translates into task performance is not extensively addressed in the literature. \cite{Choi2004} employ an experimental framework to study whether replacing one group member is associated or not with improvements in task performance. In another study, \cite{Akgun2006} use surveys to examine the interrelations between groups with a stable composition, group learning, and task performance. Additionally, \cite{VanBalen2017} empirically study how changes in the composition of groups dedicated to video game development affect the critical reception and the sales of the published games. While differing in their methods, these three studies arrive at similar conclusions: Changes in group composition do not improve the performance of already high-performing groups \citep{Akgun2006,Choi2004,VanBalen2017}. Thus, they support the \enquote{never change a winning team} strategy.

Thus, there is conflicting evidence. Prior research sometimes considers it beneficial to fix the composition of groups formed to solve complex tasks \citep{Tannenbaum2012}. In contrast, other lines of research regard adaptivity as beneficial and claim that a greater focus should be put on better understanding the temporal aspects of groups, and in particular of the adaptation of groups \citep{Bell2017,Lundin1995,Mathieu2014,Tannenbaum2012}. 
We place our research in this tension and aim at shedding additional light on these ambiguous recommendations.

\section{The Model}
\label{sec:model}

We propose an agent-based model in which agents form a group to solve a task jointly.\footnote{The model has been implemented in Python 3.7.4. The code is available \href{https://gitlab.aau.at/dablancofern/nk-model-for-dynamic-groups}{here.}} The process of forming a group formation is repeated over time, and we are particularly interested in how adapting the group more or less frequently affects the performance achieved by the group.   
The following subsections introduce the four building blocks of the model: \textit{(i)} the initialisation phase in Sec. \ref{sec:environment}, \textit{(ii)} the process of group adaptation in Sec. \ref{sec:group}, \textit{(iii)} the individual decision-making process in Sec. \ref{sec:decision}, and \textit{(iv)} the individual learning process in Sec. \ref{sec:learning}. Finally, we introduce the performance measure and the data analysis method in Sec. \ref{sec:measures}.  Table \ref{tab:variables} gives an overview of the key variables. The sequence of events during one simulation run is summarised in Fig. \ref{fig:process}. A table summarising the notation is included in Appendix \ref{app:notation}.

\begin{figure}[H]
    \centering
    \includegraphics[scale=0.6]{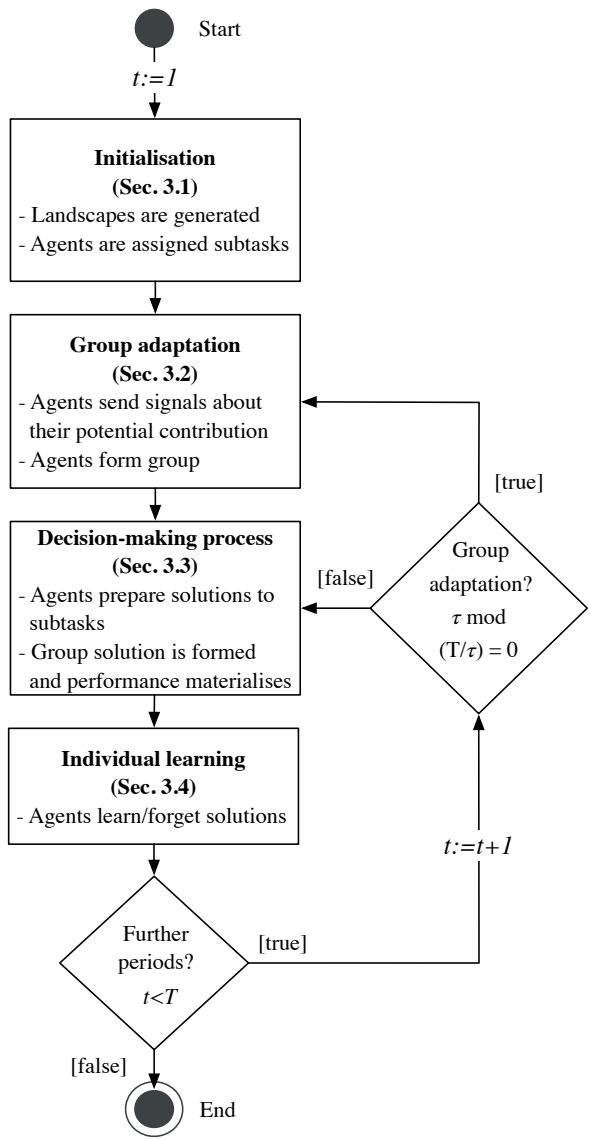}
    \caption{Sequence of events during one simulation run.}
    \label{fig:process}
\end{figure}

\subsection{Initialisation}\label{sec:environment}
\subsubsection{Task environment}\label{sec:task}

Following the \textit{NK} framework \citep{Levinthal1997}, we model the complex task as a vector $\mathbf{d}=(d_{1,}\dots, d_{N})$ of $N$ binary decisions with $K$ interdependencies among them. Each decision $d_{n} \in [0,1]$  contributes $c_{n}$ to group performance $C(\mathbf{d})$, whereby the value of performance contribution $c_{n}$ depends on decision $d_{n}$ and $K$ other decisions, following
\begin{equation}\label{eq:payoff}
    c_{n}=f(d_{n}, d_{i_1}, \dots, d_{i_K})~,
\end{equation}

\noindent where $\{i_1, \dots, i_K \} \subseteq \{1, \dots, n-1, n+1, \dots, N \}$ and $0 \leq K \leq N-1$. The contributions are randomly drawn from a uniform distribution, $c_{n}\sim U(0,1)$. The overall performance is the average of all performance contributions:
\begin{equation}\label{eq:performance}
    C(\mathbf{d})=\frac{1}{N}\sum_{n=1}^{N}c_{n}~.
\end{equation}

\noindent Since the decisions are binary, there are $2^N$ possible solutions to the complex task. We compute the performance associated with the solutions according to Eq. \ref{eq:performance} and refer to the mapping of solutions to performances as the \textit{performance landscape.} The parameter $K$ shapes the \textit{complexity} of the task and, consequently, the ruggedness of the performance landscape. If decisions are not interdependent ($K=0$), the performance landscape has a single peak. If decisions are interdependent ($K>0$), by contrast, the landscape becomes more rugged, whereby high values of $K$ result in landscapes with various local maxima \citep{Levinthal1997,Rivkin2007}.

\begin{figure}[H]
    \centering
    \includegraphics[width=1\textwidth]{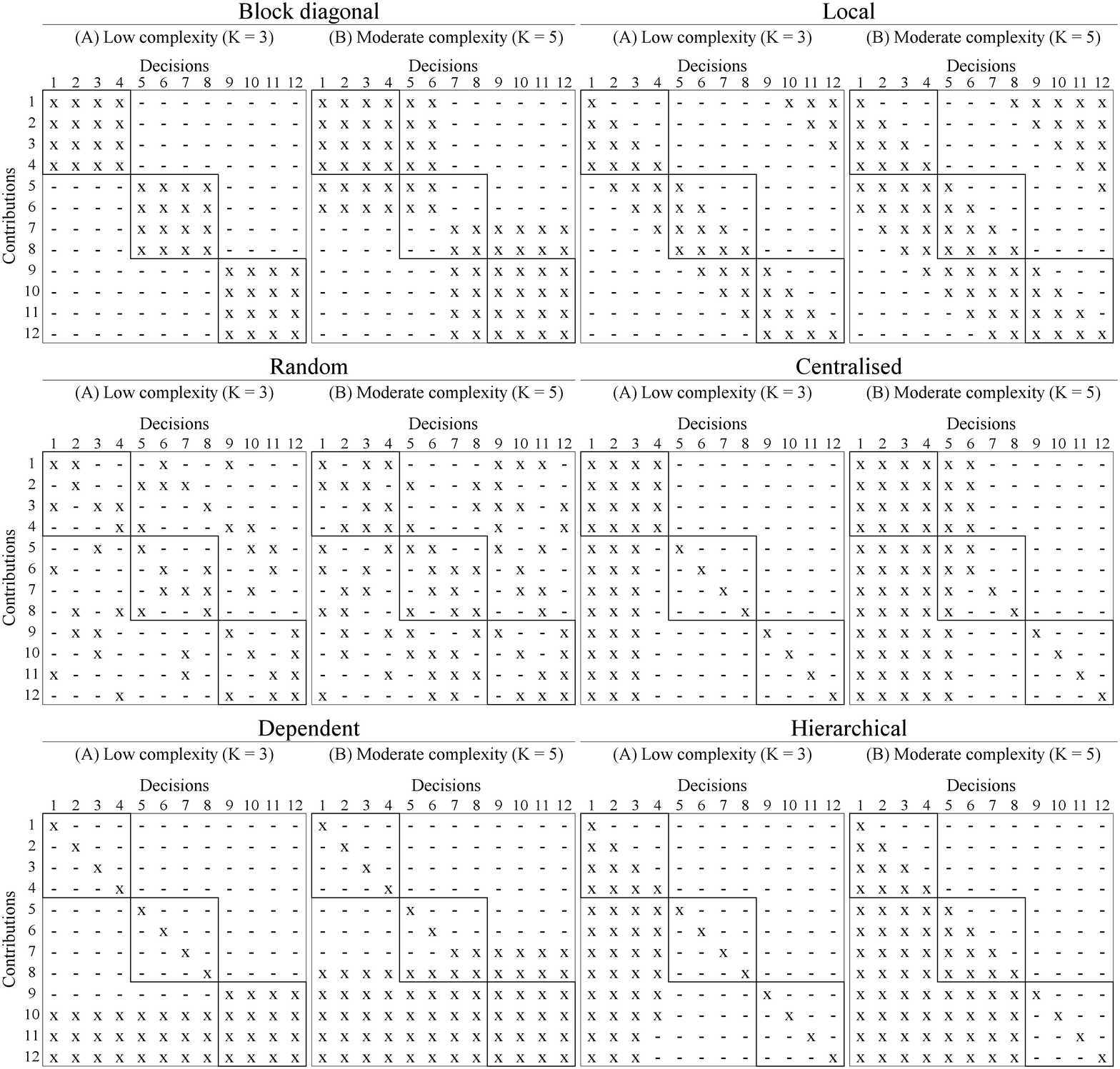}
    \captionsetup{justification=centering}
        \caption{Interdependencies between the performance contributions (represented in the y-axes) and the decisions (represented in the x-axes) are indicated with an $x$. Each contribution depends on its own decision (see Eq. \ref{eq:payoff}), so there is an $x$ in each element along the main diagonal. Solid lines indicate subtasks assigned to agents.}
    \label{fig:matrices}
\end{figure}

In this study, we model tasks of $N=12$ dimensions, and divide them into $M=3$ subtasks of equal length $S=N/M=4$. Regarding interdependencies, we consider tasks that are of either a low ($K=3$) or a moderate complexity ($K=5$) and take into account six interdependence patterns \citep[taken from][]{Rivkin2007}. Fig. \ref{fig:matrices} gives an overview of these structures; please note that the solid lines indicate the subtasks assigned to agents. 
We consider the following patterns:

\begin{itemize}
    \item \textit{Block}: Interdependencies are grouped into squares along the main diagonal. If tasks are of low complexity, they are perfectly decomposable into three independent subtasks. However, if tasks are moderately complex, slight reciprocal interdependencies between subtasks assigned to agents exist.
    \item \textit{Centralised}: Interdependencies are located within the $K+1$ first decisions. Consequently, when tasks are of low complexity, one agent highly affects the other agents' performance contributions. If tasks are moderately complex, one agent has complete, and another agent has some power to affect the other agents' performance contributions. This interdependence pattern characterises groups in which the power (to influence others) is unequally distributed. There might be reciprocal interdependencies. 
    \item \textit{Dependent}: In contrast to the centralised structure, the tasks assigned to one agent are highly dependent on the other agents' decisions. In a group setting, this interaction pattern indicates a two-stage process. At the first stage, there are two (more or less) independent tasks. At a second stage, the outcomes of these tasks are combined (and finished).
    \item \textit{Hierarchical}: The power to influence is concentrated at either one or two agents in the case of a low or a moderate complexity, respectively. In contrast to the centralised structure, there are no reciprocal interdependencies. 
    \item \textit{Local}: The interdependence patterns are organised similar to a ring structure. If tasks are of low complexity, agents only affect their neighbours, i.e., for agents one and two, the interdependencies are located below the main diagonal. At the same time, there are interdependencies between the decision assigned to agent three and the performance of agent one.
    \item \textit{Random}: Interdependencies are randomly located in the matrix. In consequence, there are interdependencies between all subtasks assigned to agents irrespective of the complexity.
\end{itemize}

\subsubsection{Agents}\label{sec:agents}

We model a population of $P$ agents who are heterogeneous concerning their \textit{capabilities}. They form a single group composed of $M<P\in \mathbb{N}$ members to solve a complex task.
Agents are bounded in their rationality \citep{Simon1957}: They want to maximise utility, but limitations in their capabilities constrain their behaviour. In particular, we limit their abilities in three ways:
\begin{enumerate}
    \item We randomly assign each agent to one of the $M=3$ areas of expertise with equal probability. This means that they are capable of solving only one of the subtasks.
    \item There are limitations in the agents' cognitive capacities. Consequently, they cannot oversee the entire solution space at a time. We denote the set of solutions known by agents at time $t$ by ${\mathbf{S}}_{mt}, =\left(\hat{\mathbf{d}}_{m1},\dots,\hat{\mathbf{d}}_{mI}\right)$

    where $\hat{\mathbf{d}}_{mi}$ represents a solution to subtask $\mathbf{d}_m$, $i=\{1,\dots,I\}\in \mathbb{N}$ and $1 \leq I \leq 2^S$. Initially, agents are not aware of the entire set of $2^S$ solutions. Instead, at $t=1$, each agent is endowed with one random solution $\hat{\mathbf{d}}_{mi}$ (i.e., $I=1$). Agents learn about solutions to their subtask over time, and $\mathbf{S}_{mt}$ grows or shrinks as a result (see Sec. \ref{sec:learning}). 
    Agents aim to optimise their immediate utility, and they do not consider the history of solutions implemented or their long-term utility when making their decisions.
\end{enumerate}

The utility of an agent assigned to the area of expertise $m$ is the weighted sum of their performance contributions $C(\mathbf{d}_{mt})$ and the performance contributions of the remaining agents of the group $C(\mathbf{d}_{rt})$, where $r=\{1,\dots,M\}\in \mathbb{N}$ and $r \neq m$. The decisions located outside the scope of the agents are the \textit{residual decisions}, and are denoted by $\mathbf{D}_{mt} = (\mathbf{d}_{1t}, \dots, \mathbf{d}_{\{m-1\}t}, \mathbf{d}_{\{m+1\}t},\dots, \mathbf{d}_{Mt})$. Agent $m$'s utility follows a linear function and is formalised in Eq. \ref{eq:utility}:

\begin{equation}\label{eq:utility}
    U \left( \mathbf{d}_{mt}, \mathbf{D}_{mt} \right) = \frac{1}{2} \cdot \left( C(\mathbf{d}_{mt}) + \cdot \frac{1}{M-1} \sum_{\substack{{r=1}\\{r\neq m}}}^{M} C(\mathbf{d}_{rt})\right) ~.
\end{equation}

\noindent Agents receive positive utility only if they are part of the group, while non-member agents receive utility equal to 0.

\subsection{Group adaptation}\label{sec:group}
A group comprises $M=3$ out of $P=30$ agents, i.e., one agent per subtask. At $t=1$, agents form the group for the first time. The objective of the group formation process is that the best-available experts join forces in a group. In this sense, a group that periodically adapts can integrate the agents with knowledge on the best-performing solutions within their ranks \citep{Savelsbergh2015}. At $t$, agents estimate the utility for each solution they know, i.e., each solution in $\mathbf{S}_{mt}$. Recall that part of their utility comes from the residual decisions. Since we omit communication with other agents, they use the residual decisions implemented at the previous period $\mathbf{D}_{m\{t-1\}}$ as a basis for their estimations. Agent $m$'s \textit{estimated utility} is then $ U \left( \mathbf{d}_{mt}, \mathbf{D}_{m\{t-1\}} \right)$.

Once the agents have made their estimations, they compute the solution in $\mathbf{S}_{mt}$ that maximises their utility at time $t$ according to
    \begin{equation}
    \label{eq:agents-decison}
    \hat{\mathbf{d}}^{*}_{mt} := \argmax_{\mathbf{d}^\prime \in \mathbf{S}_{mt}} ~ U\left(\mathbf{d}^\prime, \mathbf{D}_{m\{t-1\}} \right)~, 
    \end{equation}
\noindent and send the signal $U\left(\hat{\mathbf{d}}^{*}_{mt}, \mathbf{D}_{m\{t-1\}} \right)$. The  signal can, for example, take the form of providing letters of application, disclosing their interest in a specific task, or completing questionnaires if applicable.
The agents' signals are evaluated whenever a group is up to adapt. The agent who signals the highest estimated utility for a subtask $\mathbf{d}_m$ and, hence, appears to be most suited to fulfil the subtask is selected for the group.

Agents are assumed not to cheat and be honest and conscientious when giving signals. Also, agents always have an incentive to participate in the group since this is the only way to experience positive utility. Additionally, agents cannot observe the other agents' signals or the solutions that the other agents know, avoiding any strategic behaviour when making their estimations. Finally, we assume that all agents are fully aware of the group adaptation mechanism and its functioning.

The group adaptation process is repeated every $\tau$ period. The higher (lower) $\tau$ is, the less (more) frequently a group adapts. In this paper, we consider three different cases: 

\begin{itemize}
    \item \textit{Long-term group composition}: The group is formed only once, i.e., at $t=1$, and is stable throughout the observation period. For this case, we set $\tau=\emptyset$. 
    \item \textit{Medium-term group composition}: The group is formed at $t=1$, and the group adaptation process takes place every $\tau=10$ time steps. 
    \item \textit{Short-term group composition}: The group is formed at $t=1$ and, subsequently, adapts in every time step. For this case, we set $\tau$ equal to $1$. 
\end{itemize}

\subsection{Individual decision-making process and group strategy}\label{sec:decision}
Once agents have formed a group, they choose a particular solution to the complex task. Agents make their choices without communicating with the remaining agents of the group. This corresponds to a decentralised and distributed organisation \citep{Siggelkow2005}. 

Agents follow the decision-making rule described in Sec. \ref{sec:group}. First, agents estimate the utility $ U \left( \mathbf{d}_{mt}, \mathbf{D}_{m\{t-1\}} \right)$ for each solution they know. Next, each agent computes and implements the solution with the highest estimated utility, i.e.,  $\hat{\mathbf{d}}^{*}_{mt}$ (see Eq. \ref{eq:agents-decison}), and the group strategy at time $t$ is formed by concatenating the solutions provided by all group members:

\begin{equation}\label{eq:groupsolution}
    \mathbf{d}_{t} :=\hat{\mathbf{d}}^{*}_{1t}\frown \dots \frown \hat{\mathbf{d}}^{*}_{Mt} ~,
\end{equation}

\noindent where $^\frown$ is the concatenation of the solution to all subtasks. We calculate task performance according to Eq. \ref{eq:performance}, and the group members experience the resulting utility according to Eq. \ref{eq:utility}. Finally, all agents can observe the implemented group strategy, which serves as the basis for their decisions in the incoming period.

\subsection{Individual learning}\label{sec:learning}
Recall that we initially endow each agent with one of the $2^S=16$ solutions to their subtask. To overcome this limitation, agents have \textit{learning capabilities} that allow them to explore the solution space. Learning occurs at the end of each time step $t$. It consists of two independent mechanisms: \textit{(i)} agents \textit{discover} new solutions to their subtask, and \textit{(ii)} agents \textit{forget} solutions that are no longer relevant.

Regarding \textit{(i)}, agents discover a new solution to their assigned subtask with probability $\mathbb{P}$. The found solution differs only in one bit 
from what the agents currently know, i.e., from the elements in $\mathbf{S}_{mt}$. Thus, agents engage in a \textit{sequential exploration} of the solution space, looking for new, better-performing solutions \citep{Levinthal1997}. Regarding \textit{(ii)}, agents might forget a solution they know is not the utility-maximising at the current time step, i.e., not $\hat{\mathbf{d}}^{\ast}_{mt}$.   
Forgetting occurs with the same probability $\mathbb{P}$ as learning does. We consider probabilities in the range between $\mathbb{P}=0$ and $\mathbb{P}=1$ in steps of $0.1$.

\subsection{Parameters}\label{sec:variables}
Table \ref{tab:variables} summarises the main parameters of the model. We investigate 396 different scenarios based on the following variables: \textit{(i)} group adaptation ($\tau$), \textit{(ii)} task complexity ($K$), \textit{(iii)} the interdependence structure (see Fig. \ref{fig:matrices}), and \textit{(iv)} learning probability ($\mathbb{P}$). We observe the performance $C(\mathbf{d_t})$, as calculated in Eq. \ref{eq:performance}, and are particularly interested in how groups that adapt more or less frequently, controlled via $\tau$, affect the performance. The remaining variables are fixed during the simulations.

\begin{table}[H]
\caption{Parameters}
\label{tab:variables}
\renewcommand{\arraystretch}{1.2}
\begin{tabular}{llll}
\\ \hline
Type                                  & Variables              & Notation                    & Values                         \\ \hline
\multirow{4}{*}{Independent variables} & Task complexity        & $K$                         & \{3, 5\}                       \\
                                      & Interdependence structure & \textit{Matrix} & See Fig. \ref{fig:matrices}     \\
                                      & Group adaptation      & $\tau$                      & \{$\emptyset, 1, 10$\}         \\
                                      & Learning probability   & $\mathbb{P}$                & $\{0:0.1:1\}$ \\
                                      & Time steps             & $t$                         & $\{1:1:100\}$   \\ \hline
Dependent variable                    & Task performance       & $C(\mathbf{d_t})$           & $[0,1]$                     \\ \hline
\multirow{4}{*}{Other parameters}       
                                      & Number of decisions    & $N$                         & 12                             \\
                                      & Population of agents   & $P$                         & 30                             \\
                                      & Number of subtasks     & $M$                         & 3                              \\
                                      & Number of simulations  & $\Phi$                         & 1,500                        \\
\hline
\end{tabular}%
\end{table}

\subsection{Performance measure and analysis}\label{sec:measures}

To assure that the results are comparable across simulation runs and scenarios, we normalise the observed task performance by the maximum achievable performance. The normalised performance at each time step $t$ is computed according to $\Bar{C}_t = C\left( \mathbf{d}_{t}\right) / C^{\ast}$, where $\mathbf{d}_{t}$ stands for the solution to the task implemented at time $t$, the function $C(\mathbf{d_{t}})$ gives the corresponding performance (see Eq. \ref{eq:performance}), and $C^{\ast}$ is the maximum achievable performance in that simulation round. 

We train regression models using the normalised performances and base our analysis on the functional relationships within these models. Details on the regression models are included in Appendix \ref{app:a}. In particular, we compute partial dependencies of the performance on a subset of the independent variables included in Tab. \ref{tab:variables}. Our data analysis approach aligns with \cite{patel2018}. They, amongst others, suggest employing regressions to analyse simulation data and parameter importance and, finally, to better understand the emergence of patterns.

Let $\mathbf{X}$ be the set of all independent variables. The subset $\mathbf{X}^s$ includes either one or two independent variables that are in the scope of the analysis, and $\mathbf{X}^c$ consists of the remaining independent variables, i.e., the complementary set of $\mathbf{X}^s$ in $\mathbf{X}$. Then, $f(\mathbf{X})=f(\mathbf{X}^s,\mathbf{X}^c)$ represents the trained regression model. The partial dependence of the performance on the independent variables in scope is defined by the expectation of the performance concerning the complementary independent variables so that 

\begin{equation}
    f^s(\mathbf{X}^s)= E_c(f(\mathbf{X}^s,\mathbf{X}^c)) \approx \frac{1}{V}\sum_{i=1}^{V} f(\mathbf{X}^s,\mathbf{X}_{(i)}^c)~,
\end{equation}

 \noindent where $V$ is the number of independent variables in $\mathbf{X}^c$ and $\mathbf{X}_{(i)}^c$ is the $i^{th}$ element. By marginalising over the independent variables in $\mathbf{X}^c$, we get a function that depends only on the independent variables in $\mathbf{X}^s$.

\section{Results and Discussion}\label{sec:results}

In this paper, we aim to understand better the effects of changing a group's composition more or less frequently ($\tau$) on performance. In addition, we aim at exploring the moderating effects of \textit{(i)} learning at the individual level ($\mathbb{P}$) and \textit{(ii)} the complexity of the task at hand ($K$, \textit{Matrix}). Since we are interested in the average effects of these variables over time, we do not explicitly include time steps $t$ in as predictor in our analyses. We start by examining the overall effects of the frequency of group adaptation and the moderating factors in Sec. \ref{subsec:results-rq1}. 
Next, we separately examine the moderating factors. In particular, Sec. \ref{subsec:results-rq2} provides insights into the moderating effect of learning at the agents' level, while Sec. \ref{subsec:results-rq3} explores the moderating effect of the interdependence structures. Finally, Sec. \ref{subsec:results-rq4} considers the two factors simultaneously and examines the effects. 

\subsection{Overall effects}
\label{subsec:results-rq1}

\begin{figure}[H]
    \centering
    \includegraphics[width=0.9\textwidth]{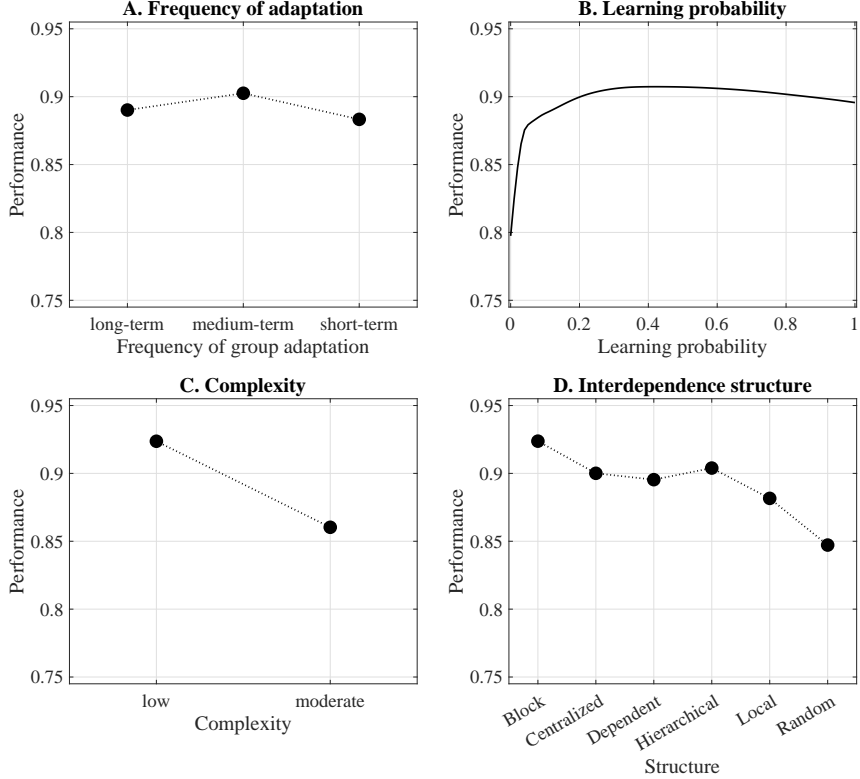}
    \caption{Overall effects.}
    \label{fig:overall-effects}
\end{figure}

We present the partial dependencies between task performance and a predictor of interest at a time, in Fig. \ref{fig:overall-effects}.

Fig. \ref{fig:overall-effects}.A shows that the overall marginal effect of group adaptation on performance is highest when groups are formed for the medium-term ($\tau=10$). However, the marginal effects of groups with a long- and short-term composition (i.e., $\tau=\emptyset$ and $1$, respectively) are only slightly smaller. 
To shed more light on the effects of group adaptation on task performance, we take a closer -- and more differentiated -- look at these results in Secs. \ref{subsec:results-rq2} to \ref{subsec:results-rq4}.

Interestingly, as Fig. \ref{fig:overall-effects}.B indicates, the functional relationship between the learning probability and task performance is more complex: There is a strong positive effect when the learning probability is relatively low. However, this effect flattens when the probability is at intermediate levels. For high learning probabilities, the slope finally turns slightly negative, enven though one might expect that task performance always increases with individual learning. Previous research argues that learning enables agents to master (challenging) task requirements, as it links consistent effort to discover efficient ways to solve a task \citep{dweck1986}. The analyses provided in Secs. \ref{subsec:results-rq2} and \ref{subsec:results-rq4} take a closer look at the effect of learning in the relationship between group adaptation and task performance.

Regarding task complexity, Fig. \ref{fig:overall-effects}.C shows that task performance decreases as complexity increases. Finally, Fig. \ref{fig:overall-effects}.D suggests a relationship between the structure of interdependencies and task performance. In particular, the more pronounced cross-interdependencies between agents are, the lower task performance is. These findings are in line with previous research. \cite{Levinthal1997}, \cite{Rivkin2000}, and \cite{leitner14}, for example, argue that task complexity and the interdependence structure affect the performance landscape's ruggedness. Since the global maximum is more difficult (easier) to find on rugged (single peaked) landscapes, complexity is negatively correlated with task performance. Sections \ref{subsec:results-rq3} and \ref{subsec:results-rq4} provide a more detailed analysis of this relationship.

\subsection{Group adaptation and the moderating effect of the learning probability}
\label{subsec:results-rq2}

\begin{figure}[H]
    \centering
    \includegraphics[width=1\textwidth]{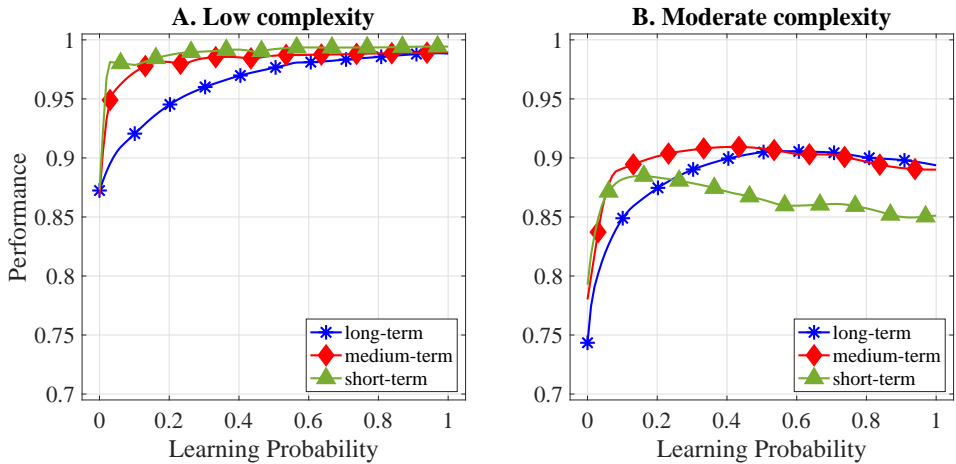}
    \caption{Partial dependencies between task performance and the learning probability.}
    \label{fig:learning}
\end{figure}

This section analyses the overall effects presented in Sec. \ref{subsec:results-rq1} in more depth by exploring the effects of learning at the agent's level. Therefore, we vary the probability of individual learning $\mathbb{P}$ between $0$ and $1$, consider tasks of low or moderate complexity, and use block interdependence structures for all cases. The partial dependencies for the analysed scenarios are plotted in Figs. \ref{fig:learning}.A and \ref{fig:learning}.B for tasks of low and moderate complexity, respectively. 

The results indicate that there is indeed a moderating effect of learning at the agents' level. For tasks of low complexity, the performance increases as agents learn with a higher probability. Groups of a medium- and short-term composition achieve similar performances for all learning probabilities and higher performances than groups that do not adapt. For these groups, at relatively low levels of learning, task performance reacts more strongly to an increase in the learning probability the more frequently they adapt. Still, there are hardly any increases in performance beyond the threshold of $0.1$. The results indicate that groups of a long-term composition can compensate for the lower performance because of a less frequent group adaptation by learning. Consequently, the performances converge to almost the same level for all three frequencies of group adaptation and high learning probabilities. 

The results follow a different pattern for scenarios with moderately complex tasks (see Fig. \ref{fig:learning}.B).  Recall, for low complexity tasks, encouraging agents to learn never harms performance. In contrast, for moderately complex tasks, too much learning might result in marginal adverse effects. The tipping point, i.e., the learning probability at which the marginal effects of learning turn negative, depends on the frequency of group adaptation. In particular, the more frequently groups adapt, the lower the learning probability at the tipping point. As a consequence of marginal adverse effects, groups of a short-term composition are worst off when the learning probability is higher than $\mathbb{P} \approx 0.25$. 

Previous research has ambiguous results regarding the effects of learning in groups and its interrelation with group adaptation. For example, \cite{Choi2005} and \cite{Perretti2007} argue that group adaptation fosters creativity in problem-solving and increases in performance. Their findings align with the argumentation provided in \cite{dweck1986}. She argues that learning assures a consistent effort to find efficient ways to solve a task. The findings also in line with the resource based view of the firm \citep{hamel1993}, according to which long-term investments into competencies, i.e., learning and long-term groups in our case, sustain well-performing groups and, in consequence, competitive advantage. We show that this argumentation holds true for tasks of low complexity. However, medium- or short-term composition groups do not perform comparatively worse (see Fig. \ref{fig:learning}.A). This might be the case because tasks are of low complexity. In such cases, the learning outcomes might be \enquote*{enabling capabilities} \citep{Leonard1998} or \enquote*{table stakes} \citep{hamel1994}, i.e., capabilities that can be easily transferred, copied, and are not sufficient in themselves to comparatively distinguish a group from other groups.

In contrast, \cite{Bartsch2013} and \cite{Sergeeva2018} find that group adaptation might offset the positive effect of learning on performance. We shed additional light on these conflicting results.  We show that creating competencies to solve tasks by promoting learning can be complemented by attracting new and competent group members \citep{holland2007}, as long as group adaptation does not take place too often. This is in line with the resource-based view of the firm since creating and retaining competencies results in relatively higher performances. However, suppose a group faces a task of moderate complexity. In that case the findings presented in \cite{Bartsch2013}, \cite{Sergeeva2018}, and \cite{bunderson2003} hold true: Groups that are high on learning might overemphasise exploration at the cost of performance. 

\subsection{Group adaptation and the moderating effect of the interdependence structure} 
\label{subsec:results-rq3}

\begin{figure}[H]
    \centering     \includegraphics[width=\textwidth]{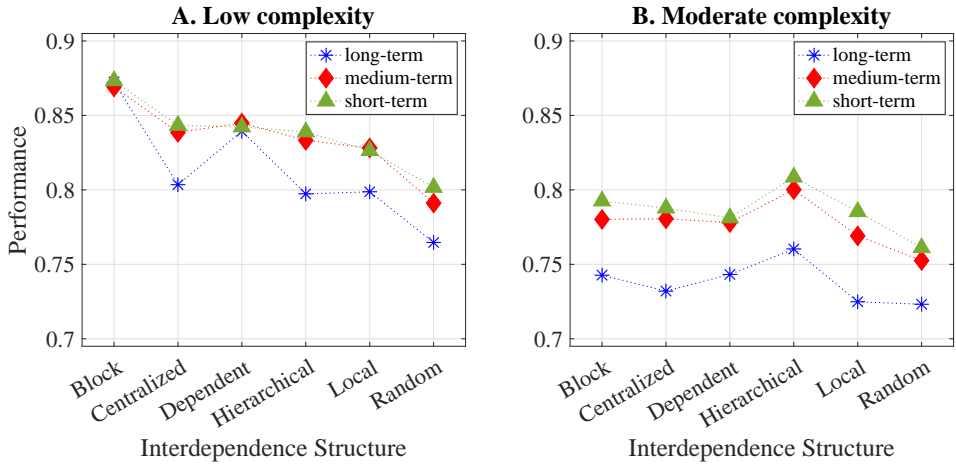}
    \caption{Partial dependencies between task performance and interdependence structures.}
    \label{fig:structure}
\end{figure}

This section analyses the overall effects presented in Sec. \ref{subsec:results-rq1} in more detail, by adding variations in the interdependence structure (see Fig. \ref{fig:matrices}). Again, we vary the frequency $\tau = \{\emptyset,1,10\}$ at which groups adapt. We fix the individual learning probability to $\mathbb{P}=0$. The partial dependencies for scenarios considering tasks of low complexity ($K=3$) are plotted in Fig. \ref{fig:structure}.A, while the results for moderately complex ($K=5$) tasks are presented in Fig. \ref{fig:structure}.B.

When task complexity is low, the results show that the interdependence structure indeed has a moderating effect. In particular, the results indicate that the extent to which agents experience an influence from other agents appears to matter. In the case of a block structure, there are no cross-interdependencies, the performance is the highest, and there are no differences in performance between groups. 
When there is a dependent structure, \textit{one} agent is influenced from outside their area of responsibility. The performances are lower than in the case of a block structure, and there are again no differences between groups. 
In the cases of a centralised, hierarchical or local structure (see Fig. \ref{fig:matrices}), \textit{two} agents are influenced by the other agents' decisions. In these cases, the following moderating effect unfolds: Groups that adapt ($\tau = 1$ or $10$) are better off than groups that do not adapt ($\tau=\emptyset$). It is also worth noting that it is almost negligible how often groups adapt since similar partial dependencies can be observed for frequent and moderate adaptations. In the case of a random interdependence structure (see Fig. \ref{fig:matrices}), all \textit{three} agents are influenced by the other agents' decisions. Again, we observe relatively weak effects for groups with a long-term composition. The partial dependence between task performance and the structure is highest for groups with a short-term composition. \cite{Rivkin2000} and \cite{maccormack2012}, for example, argue that self-contained structures (the block pattern in our case) result in higher task performances, while interdependencies between subtasks come at the cost of performance. For low complexity tasks, we show that their argumentation holds true, whereby it appears to matter how many agents experience influence from the other agents. 

Recall that agents send signals about their estimations of what they can contribute to the group solution (see Sec. \ref{sec:group}). If agents experience no or only minor influence from outside their area of responsibility, they can predict their contribution quite well. Consequently, it is assured that the most capable agents join forces in the group formed initially. Thus, there is no reason for any further adaptations with a high precision of the agents' predictions. However, if cross-interdependencies between agents exist, the accuracy of the agents' predictions decreases because of behavioural uncertainty. In consequence, there is a chance that \textit{not} the most capable agents join forces when forming a group for the first time. Thus, whenever the precision of the agents' predictions is low, repeated adaptations increase the chances that the most capable agents join forces in a group. This argumentation is in line with the findings presented in \cite{Licalzi2012}, \cite{buyukboyaci2019}, and \cite{eppler2000}. They argue that groups need to be heterogeneous to solve tasks efficiently, and the group  members' skills need to complement each other. In our model, we observe how groups \textit{emerge} so that finally, those individuals who are best prepared for the task, join the group. A more frequent group adaptation indicates more efforts to search for group members with complementary skills. In consequence, the performance is relatively higher.

In Fig. \ref{fig:structure}.B, we plot the partial dependencies for moderately complex tasks. Due to the complexity inherent to the task, agents almost always experience interdependencies from outside their area of responsibility. Consequently, they struggle with making precise predictions of their potential contributions to the overall solution. Therefore, adaptive groups are better off in all cases, which is in line with the argumentation provided in the previous paragraph. Still, the frequency at which groups adapt has a minor impact on task performance. It is worth noting that group adaptation leads to comparatively higher performances only in cases with relatively pronounced interdependencies between agents. This includes the centralised, hierarchical, local, and random structure in Fig. \ref{fig:structure}.A and all cases included in Fig. \ref{fig:structure}.B. These findings align with the arguments put forward by \cite{Choi2005} and \cite{Perretti2007}. They argue that group adaptation positively affects performance because it fosters creativity. Thus, we show that these arguments only hold true for relatively complex situations.

\subsection{Group adaptation and simultaneous moderating effects}
\label{subsec:results-rq4}

\begin{figure}[H]
    \centering
    \includegraphics[width=0.95\textheight, angle=90]{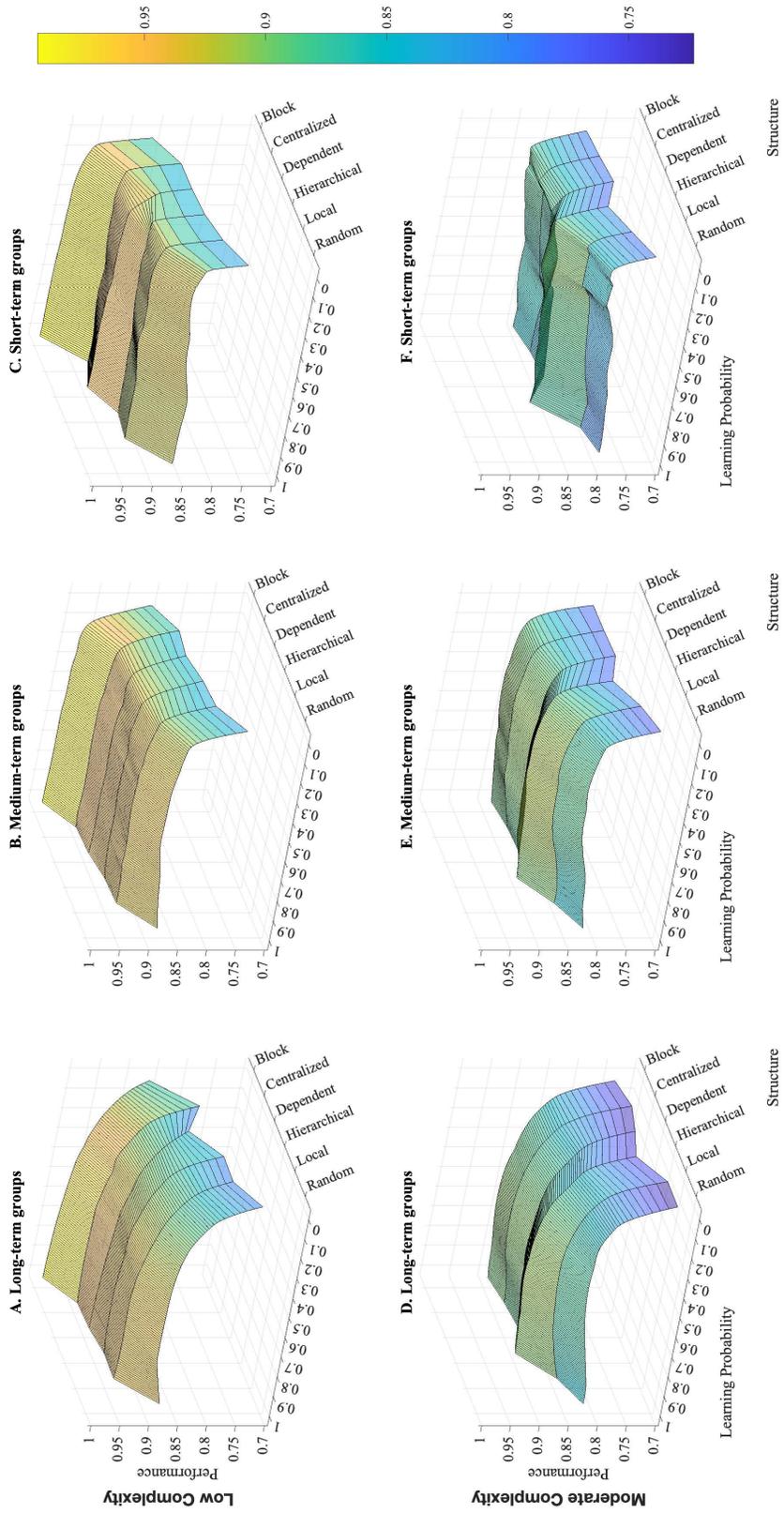}
    \caption{Partial dependencies for a simultaneous variation of moderating factors.}
    \label{fig:simul}
\end{figure}

The analysis in this section considers a simultaneous variation in the learning probability and the interdependence structure and studies the resulting moderating effects. In Fig. \ref{fig:simul}, we plot the partial dependencies for each frequency of group adaptation and tasks of either a low or moderate complexity.

Fig. \ref{fig:simul}.A--C,  presents the results for tasks of low complexity. In line with the results for block interdependence structures presented in Sec. \ref{subsec:results-rq2}, increasing the learning probability up to $0.1$ leads to an increase in performance. The more frequently groups adapt, the more strongly the performance reacts to increases in the learning probability at relatively low levels.

For the case of groups with a long- and medium-term composition and learning probabilities beyond $0.1$, the marginal effects become less pronounced. Also, we observe that the performance reacts most strongly (weakly) to learning in the case of a block (random) structure, whereas the performances for the remaining structures converge to almost the same level. It is worth noting that the relative advantage of the dependent structure in the case of long-term groups and no learning (see Fig. \ref{fig:simul}.A) disappears with higher learning probabilities. Thus, most of the results for groups with a long- and medium-term composition and tasks of low complexity presented in Secs. \ref{subsec:results-rq2} and \ref{subsec:results-rq3} are robust against simultaneous variations in the learning probability and the interdependence structure. 

On the contrary, high learning probabilities for groups with a short-term composition might result in marginal adverse effects. This is particularly the case for tasks with a centralised and hierarchical interdependence structure. Again, we observe the highest (lowest) performance for a block (random) interdependence structure. For the remaining interdependence structures, we observe that the effect of learning on performance stabilises beyond a learning probability of around $0.1$. 

The results for tasks of moderate complexity are presented in Fig. \ref{fig:simul}.D--F. Recall, in Sec. \ref{subsec:results-rq2} we focused on block interdependence structures and found that learning substantially affects performance when groups adapt. The slopes of the surfaces plotted in Fig. \ref{fig:simul}.D--F (in the range of a learning probability between $0$ and $0.2$) indicate that this finding is robust across all interdependence structures. Section \ref{subsec:results-rq2} also found that increasing the learning probability beyond a certain threshold (tipping point) might unfold marginal adverse effects. Also, the tipping points are contingent on the frequency of group adaptation. The results for moderately complex tasks presented in Fig. \ref{fig:simul} confirm this finding for most interdependence structures. 
Only when interdependencies follow a hierarchical pattern, there appear to be no such (or less pronounced) marginal adverse effects. This might be the case because interdependencies between agents are less pronounced for moderately complex tasks in this structure. 

For the remaining interdependence structures, we see that the tipping points are at a learning probability of around $0.1$ in Fig. \ref{fig:simul} F, and move to higher learning probabilities for groups of a medium- and a long-term composition (see Figs. \ref{fig:simul}.D and E). 
Thus, we show that the performance is correlated to the extent to which interdependencies \textit{between} agents exist, which is in line with previous research on efficient task allocation \citep{Rivkin2000,maccormack2012}. However, suppose groups are high on innovation (learning and adaptation). In this case, cross-interdependencies between agents appear to pose cognitive requirements on agents that come at the cost of learning efficiency and performance, which is in line with the (intrinsic) cognitive load theory \citep{sweller2011}. In consequence, in Fig. \ref{fig:simul}.D--F, we observe that the performances achieved by groups of a short-term composition are lower than those achieved by the other groups for high learning probabilities. Recall, in Sec. \ref{subsec:results-rq3} we found that, for the case of no learning, long-term groups are worse off than groups of a medium- or short-term composition. This pattern shifts as we increase the learning probability to moderate or high levels.

\section{Summary and Conclusions}\label{sec:conclusion}

In this paper, we have implemented an agent-based model to analyse the effect of group adaptation on task performance. Our results suggest that changing a winning team might indeed unfold positive effects on performance; but we show that a more differentiated approach is required to manage groups efficiently. Overall, groups are always well-advised to adapt if they are low on learning. However, if groups are already high on learning, adaptation might not always be a sustainable strategy since adverse effects on performance are lurking. The task's interdependence pattern and complexity reinforce both positive and negative impacts of learning on performance. Our analysis helps explain the ambiguous results of previous research by taking learning, complexity, and the interdependence structure into account.

Our research is, however, not without its limitations. We consider human decision-makers who have perfect foresight when evaluating their performance landscape \citep{Hendry2002}, do not suffer from decision-making biases \citep{Kahneman1982}, and can handle tasks of any complexity \citep{Pennington1986}. Moreover, we omit communication between agents. Also, future research might include coordination mechanisms \citep{Rivkin2003}. Finally, we assume that the competencies developed by individual learning can easily be transferred between groups. Future research could consider skills that are (non-transferable) core competencies \citep{hamel1993}.

\subsection*{Availability of data and code}
The code of the model is available \href{https://gitlab.aau.at/dablancofern/nk-model-for-dynamic-groups}{here.} Simulation data and the code for the analysis are provided  \href{https://gitlab.aau.at/dablancofern/nk-model-for-dynamic-groups/-/tree/data}{here.}

\newpage
\appendix

\section{Notation}
\label{app:notation}

\begin{table}[H]
\caption{Notation}
\label{tab:notation}
\renewcommand{\arraystretch}{1.2}
\begin{tabular}{p{0.15\textwidth}p{0.75\textwidth}}
\\ \hline
Notation & Description \\
\hline
$N$                 &   Number of decisions \\
$K$                 &   Number of interdependencies between decisions \\
$M$                 &   Number of subtasks / agents in the group \\
$P$                 &   Total number of agents \\
$I$                 &   Number of solutions known by an agent \\
$C(\cdot)$          &   Performance function \\
$U(\cdot)$          &   Utility function \\
$\frown$            & Concatenation operator\\
$\tau$              &   Periods between two rounds of group adaptation\\
$\mathbb{P}$        &   Learning probability\\    
$t$                 &   Time steps\\
$\mathbf{d}$        &   Vector of binary decisions \\
$d_n$               &   Decision $n$\\
$c_n$               &   Performance contribution of decision $n$\\
$\overline{C}_t$    &   Normalized performance in $t$\\
${C}^\ast$          &   Maximum achievable performance\\
$\mathbf{d}_t$      &   Solution to the entire task implemented in $t$ \\
$\mathbf{S}_{mt}$   &   All solutions known by agent $m$ to solve the assigned subtask at time $t$ \\
$\hat{\mathbf{d}}_{mi}$&   $i^{th}$ solution known by agent $m$ to solve the assigned subtask \\
$\hat{\mathbf{d}}_{mt}^{\ast}$&   Utility-maximizing solution known by agent $m$ to solve the assigned subtasks in $t$\\
${\mathbf{d}}_{mt}$ &   Solution implemented by agent $m$ to solve their subtask in $t$\\
${\mathbf{D}}_{mt}$ &   Solutions implemented by agents other than $m$ to solve their subtask in $t$\\
${\mathbf{X}}$ &   Set of independent variables\\
${\mathbf{X}}^s$ &   Set of independent variables in scope of the analysis\\
${\mathbf{X}}^c$ &   Complementary set of independent variables\\
\hline
\end{tabular}%

\end{table}

\section{Data analysis}
\label{app:a}

We use simulated data to train regression neural networks and analyze the data following the procedure introduced in Sec. \ref{sec:measures}. Table \ref{tab:app} gives an overview of the used datasets and the corresponding parameter settings, the trained models (type neural network used in the regression and number of nodes), and RMSE and R$^2$. Since we are interested in the average effects, all  models include data for periods $t=\{1,\dots,100\}$. For details about the parameters included in the trained regression model, see Sec. \ref{sec:variables} and Tab. \ref{tab:variables}. 

We trained the regression neural networks in the Matlab Regression Learner App, which returns a number of models for every dataset. Out of the models, we selected those with the lowest RMSE (root-mean-square error) and highest R$^2$ for our analysis.

\begin{table}[H]
\caption{Regression analyses}
\label{tab:app}
\renewcommand{\arraystretch}{1.2}
\begin{tabular}{@{\extracolsep{4pt}}llllllll@{}}
\\\hline
\multicolumn{4}{c}{Parameters} & \multicolumn{2}{c}{Neural Network} & 
\multicolumn{2}{c}{Validation}\\
\cline{1-4} 
\cline{5-6}
\cline{7-8}
$K$ &   \textit{Matrix}  &  $\tau$   & $\mathbb{P}$  &  Type & Nodes & RMSE    & R$^2$ \\ 
\hline
\multicolumn{2}{@{} l}{\textbf{Section \ref{subsec:results-rq1}:}}\\
\{3,5\}   & all       & $\emptyset$   &$\{0:0.1:1\}$ & Wide       & 100       & $0.0050$  & $0.99$\\ 
\hline
\multicolumn{2}{@{} l}{\textbf{Section \ref{subsec:results-rq2}:}}\\
3         & Block     & $\emptyset$   &$\{0:0.1:1\}$ & Wide     & 100   & $0.0017$  & $1$\\
5         & Block     & $\emptyset$   &$\{0:0.1:1\}$ & Wide     & 100   & $0.0018$  & $1$\\
3         & Block     & $10$          &$\{0:0.1:1\}$ & Wide     & 100   & $0.0047$  & $1$\\
5         & Block     & $10$          &$\{0:0.1:1\}$ & Wide     & 100   & $0.0030$  & $1$\\ 
3         & Block     & $1$           &$\{0:0.1:1\}$ & Wide     & 100   & $0.0024$  & $1$\\
5         & Block     & $1$           &$\{0:0.1:1\}$ & Wide     & 100   & $0.0020$  & $1$\\ 
\hline
\multicolumn{2}{@{} l}{\textbf{Section \ref{subsec:results-rq3}:}}\\
3         & all       & $\emptyset$   &$0$ & Wide       & 100       & $0.0136$  & $0.87$\\
5         & all       & $\emptyset$   &$0$ & Tri-layered & 10-10-10  & $0.0073$  & $0.79$\\ 
3         & all       & $10$          &$0$ & Wide       & 100       & $0.0139$  & $0.78$\\
5         & all       & $10$          &$0$ & Tri-layered & 10-10-10  & $0.0094$  & $0.86$\\  
3         & all       & $1$           &$0$ & Wide       & 100       & $0.0129$  & $0.78$\\
5         & all       & $1$           &$0$ & Tri-layered & 10-10-10  & $0.0062$  & $0.91$\\  
\hline
\multicolumn{2}{@{} l}{\textbf{Section \ref{subsec:results-rq4}:}}\\
3         & all       & $\emptyset$   &$\{0:0.1:1\}$ & Wide       & 100       & $0.0025$  & $1$\\
5         & all       & $\emptyset$   &$\{0:0.1:1\}$ & Wide       & 100       & $0.0028$  & $1$\\
3         & all       & $10$          &$\{0:0.1:1\}$ & Wide       & 100       & $0.0052$  & $0.99$\\
5         & all       & $10$          &$\{0:0.1:1\}$ & Wide       & 100       & $0.0039$  & $1$\\ 
3         & all       & $1$           &$\{0:0.1:1\}$ & Wide       & 100       & $0.0033$  & $1$\\
5         & all       & $1$           &$\{0:0.1:1\}$ & Wide       & 100       & $0.0031$  & $1$\\
\hline
\end{tabular}%
\end{table}

\newpage
\bibliographystyle{apacite}
\bibliography{biblio}

\end{document}